\documentclass [aps,twocolumn,superscriptaddress,altaffilletter,lengthcheck,tightenlines,showpacs,showkeys]{revtex4}

\newcommand{\va}{\varphi}
\newcommand{\al}{\alpha}

\newcommand{\ben}{\begin{eqnarray}}
\newcommand{\een}{\end{eqnarray}}
\newcommand{\be}{\begin{equation}}
\newcommand{\ee}{\end{equation}}
\newcommand{\ba}{\begin{eqnarray}}
\newcommand{\ea}{\end{eqnarray}}
\newcommand{\n}{\label}
\newcommand{\no}{\noindent}
\newcommand{\la}{\lambda}
\newcommand{\ga}{\gamma}
\newcommand{\ro}{\rho}

\usepackage[dvipdf]{epsfig}
\usepackage{color}
\begin{document}


\title{Tachyon and Quintessence in Brane-Worlds}


\author{Luis P. Chimento}\email{chimento@df.uba.ar}
\affiliation{ Departamento de F\'{\i}sica, Facultad de Ciencias Exactas y
Naturales,  Universidad de Buenos Aires, Ciudad
Universitaria, Pabell\'on I, 1428 Buenos Aires, Argentina}
\author{M\'onica Forte }\email{forte.monica@gmail.com}
\affiliation{ Departamento de F\'{\i}sica, Facultad de Ciencias Exactas y
Naturales,  Universidad de Buenos Aires, Ciudad
Universitaria, Pabell\'on I, 1428 Buenos Aires, Argentina}
\author{Gilberto M. Kremer}\email{kremer@fisica.ufpr.br}
\affiliation{ Departamento de F\'\i sica,
Universidade Federal do Paran\'a, Caixa Postal 19044, 81531-990 Curitiba, Brazil}
\author{Mart\'{\i}n G. Richarte}\email{martin@df.uba.ar}
\affiliation{ Departamento de F\'{\i}sica, Facultad de Ciencias Exactas y
Naturales,  Universidad de Buenos Aires, Ciudad
Universitaria, Pabell\'on I, 1428 Buenos Aires, Argentina}

\begin{abstract}
Using tachyon or quintessence fields along with a barotropic fluid on the brane we examine the different cosmological stages in a Friedmann-Robertson-Walker(FRW) universe, from the first radiation scenario to the later era dominated by cosmic string networks. We introduce a new algorithm to generalize previous works on exact solutions and apply it to study tachyon and quintessence fields localized on the brane. We also explore the low and high energy regimes of the solutions. We show that the tachyon and quintessence fields are driven by an inverse power law potential. Finally, we find several simples exacts solutions for tachyon and/or quintessence fields.
\end{abstract}

\date{\today}



\maketitle


\section{Introduction}


Recent progress in the superstring theory has shown that the strongly coupled
$E_8\times E_8$ heterotic string can be identified as the 11-dimensional
limit of M- theory compactified on an $S^1/Z_2$ orbifold with a set of $E_8$
gauge fields at each ten-dimensional orbifold fixed plane~\cite{hw1,hw2}.
Furthermore, there exists a consistent compactification of this M-theory limit
on a Calabi-Yau threefold, such that for energies below the unification scale
there is a regime where the Universe appears five-dimensionally. This
five-dimensional regime represents a new setting for early universe cosmology,
which has been traditionally studied in the framework of the four-dimensional
effective action. The theory {is developed} in a five-dimensional space which is a
product of a smooth four-dimensional manifold times the orbifold $S^1/Z_2$.
Therein the matter fields are confined to the four-dimensional spacetime while
gravity can propagate in the full spacetime.
This model and its noncompact analogs
\cite{Gogberashvili:1999ad,Ra1,Cohen:1999ia,Gregory:1999gv}
(see Ref.\ \cite{Visser:1985qm} for an account of earlier works)
provide a novel setting for discussing various conceptual and
phenomenological issues related to compactification of extra dimensions
in models motivated by the M- theory.

Two interesting possibilities were suggested in \cite{Ra1,Ra2}. In the
two-brane model \cite{Ra2}, the  branes have tensions of {opposite}
sign and the bulk cosmological constant is chosen in such a way that the
classical solution describes five-dimensional space-time whose
four-dimensional slices are flat.

It was shown that mass scales on the
negative tension brane can be severely suppressed, leading to a solution of the
hierarchy problem. Of course, this assumes that we live on the negative
tension brane. It was shown by Shiromizu, Maeda, and Sasaki \cite{Sh1} that
the effective Einstein field equations on the negative tension brane involve a
negative gravitational constant which means that gravity would be repulsive
instead of attractive. However, they showed that one does recover the correct
Einstein equations in the low energy limit on the positive tension brane. More
recently, it has been shown \cite{Cs2} that the problem with the negative
tension brane may disappear if the extra dimension is stabilized by a radion
field.

In the second scenario we live on the positive
tension brane and the negative tension brane is moved off to infinity. Thus,
in this scenario the extra dimension is infinite in extent
and there is a single gravitational bound state confined to the
brane that corresponds to the graviton. Even though the
extra dimension is infinite, the effective gravitational interaction on the
brane is that of a four-dimensional space-time with some very small
corrections.

Randall and Sundrum  have suggested\cite{Ra2}, \cite{rs2} that four-dimensional  gravity
may be recovered in the presence of an infinite fifth dimension provided that
we live on a domain wall embedded in anti--de Sitter space. Their linearized
analysis showed that there is a massless bound state of the graviton
associated with such a wall as well as a continuum of massive Kaluza-Klein
modes. More recently, linearized analyses have examined the space-time produced
by matter on the domain wall and concluded that it is in close agreement with
four-dimensional Einstein gravity \cite{gt,gkr}. In Ref. \cite{Ra1} a new
static solution to the $5$-D Einstein equations was presented  in which
spacetime is flat on a 3-brane with positive tension provided that the bulk
has an appropriate negative cosmological constant . Even if the fifth
dimension is uncompactified, standard $4$-D gravity (specifically, Newton's
force law) is reproduced on the brane. In contrast to the compactified case
\cite{ADD}, this follows because the near-brane geometry traps the massless
graviton.

A natural extension of the Randall-Sundrum model is to
include higher-order curvature invariant in the bulk action. This kind of terms arise in the
anti--de Sitter/ conformal field theory correspondence as next to  leading order corrections to the conformal field theory \cite{cftbrane}.
The Gauss-Bonnet combination of curvature invariants is quite relevant in five dimensions because it is the only invariant which leads to field equations of second order linear in the highest derivative thereby ensuring a unique solution\cite{egb5d}. The Einstein-Gauss-Bonnet equations projected onto the brane lead to a complicated Hubble equation in
general (see \cite{charmoux}, \cite{kmaeda}, \cite{dufaux}, \cite{egbotros}). However, it reduces to a very simple equation $H^2 \approx \rho^{\theta}$ with $\theta = 1, 2, 2/3$ in limiting cases corresponding to general relativity, Randall-Sundrum and Gauss--Bonnet regimes, respectively.

An interesting model that incorporates modification of gravitational laws at large distances was proposed by Dvali, Gabadadze and Porrati in \cite{dgp}. The model describes a brane with four-dimensional worldvolume, embedded into flat five-dimensional bulk. Ordinary matter is supposed to be localized on the brane, while gravity can propagate in the bulk. A crucial ingredient of
the model is the induced Einstein-Hilbert action on the brane, because it allows under a non trivial mechanism, to recover the four-dimensional Einstein gravity at moderate scales \cite{dgpcosmo}.


In recent times a great amount of work has been invested in studying the inflationary
model with a tachyon field. The tachyon field associated with unstable D-branes might be
responsible for cosmological inflation in the early evolution of the Universe, due to tachyon
condensation near the top of the effective scalar potential \cite{sentachyon}, which could also add some new form of cosmological dark matter at late times \cite{samit}. Cosmological implications of this rolling tachyon were first studied by Gibbons \cite{tachyongibbons} who showed that it is quite natural to consider scenarios in which inflation is driven by the rolling tachyon.
Later,  an extended version of the tachyon Lagrangian \cite{extendedtachyon} was found and it was shown how this new version could be related to the generalized Chaplygin gas.
In addition, using  a phenomenological  approach was described
which conditions the potential and the tachyon mass must fulfill in order
to provide enough $e$-folds for inflation and to have density perturbations of the correct
magnitude \cite{bruselas}. In addition, the structural stability of tachyonic inflation against
changes in the shape of the potential  was carefully explored in Ref. \cite{ruthtachyon}.

One of the main elements within the traditional brane worlds is a quadratic correction  in the energy-momentum tensor projected onto  the brane. Basically, this means that the Einstein's equations on the brane  turn out  to be a highly complex nonlinear system. Then, it is a very hard task to find exact solutions for  brane worlds. In order to gain a better insight about the physics underling in the brane models, some authors  have found particular cosmological solutions. Recently, for a tachyon field two solution have been reported for the inflationary era on the brane. Basically, it was found that compared to scalar field inflation, tachyonic inflation is not smooth \cite{Paultachyoninflation}. Also, the power law solution was obtained for inverse power potential in the Dvali, Gabadadze and Porrati model with tachyon source \cite{tachyonpl}. Others aspects concerning the inflation with tachyon source, in the brane cosmologies,  were studied. For example, in  Ref. \cite{tachyonherrera} the {Chaplygin} equation of state is used to model {the} tachyonic matter on the brane.

Maartens et al \cite{MaartensExact} studied brane-world universe with massive scalar field. They focus on the effects of the chaotic inflationary models when the kinetic energy is much less than the potential energy ( $V\gg\dot{\phi}^{2}$). Later, the opposite case ($V\ll\dot{\phi}^{2}$)  was explored in Ref.\cite{PaulExact}; this situation leads to the stiff matter  equation of state ($p_s=\rho_s$) for the scalar field.
Finally, we want to mention that Hawkins and Lidsey \cite{HawkinsLidsey} proposed different algorithms to find exacts solutions on brane {models}.

The aim of this paper is to develop a general method to integrate brane--world cosmological theories
containing scalar  and tachyon fields with a self-interaction potential.  We assume that the energy density of the matter and scalar (or tachyon)
fields are functions of the scale factor. We have found that it is possible to reconstruct the
scalar (or tachyon) field potential for simple cosmological solutions. We also examine the high and low energy limits of the solutions. Some of the solutions are used to describe the cosmological evolution in the inflationary, radiation,  matter and cosmic string eras.


\section{The Theory}


Let us consider a 5D space-time with a brane world located at $\Phi(X^{a})=0$, where $X^{a}$, $a=0,1,2,3,4$ are five-
dimensional coordinates. The effective action in five-dimensional manifold is given by (see \cite{Sh1}, \cite{Maartens}, \cite{MaWa})
\ben\nonumber
S=\int d^{5}{X}\sqrt{-g_{5}}\left(\frac{1}{2k^{2}_{5}}R_{5}-\Lambda_{5}\right)\\
+ \int_{\Phi=0}d^{4} x\sqrt{-g_{4}}\left(\frac{1}{k^{2}_{5}}K^{\pm}-\lambda + L^{matter}\right),
\een
with $k^{2}_{5}=8\pi G_{5}$ is the five-dimensional gravitational coupling constant and where $x^{\mu}$
are the induced four-dimensional brane-world coordinates, so $\mu$ takes values $\{0,..,3\}$. $R_{5}$ is the 5D intrinsic curvature in the bulk, $K_{\pm}$
is the intrinsic curvature on either side of the brane and $\lambda$ is the brane tension. Here, $L^{matter}$ corresponds to the Lagrangian for the matter fields.

On the five-dimensional space-time (usually named as the bulk), with the negative vacuum energy $\Lambda_{5}$ and the  brane energy momentum as the source of the gravitational field, the Einstein field equations are given by
\begin{equation}
G_{ab}=k^{2}_{5}T_{ab},~~ T_{ab}=-\Lambda_{5}g_{ab} + \delta(Y)\left( -\lambda + T^{matter}_{ab}\right).
\end{equation}
In this space-time the brane is a fixed point of the ${Z}_{2}$ symmetry. In the following Latin indices run from $0$ to $4$ while  Greek indices take the values $\{0,..,3\}$. Assuming a metric of the form $ds^{2}=(n_{a}n_{b }+ h_{ab})dx^{a}dx^{b}$, with $n_{a}dx^{a}=d\chi$ the unit normal to the $\chi=\mbox{cte}$ hyper-surfaces and $h_{ab}$ the induced metric on $\chi=\mbox{cte}$, the effective four-dimensional gravitational equations on the brane (which can be deduced from  the Gauss-Codazzi equations) take the form \cite{Sh1}:
\begin{equation}
G_{\mu\nu}=-\Lambda g_{\mu\nu} + k^{2}_{4}T_{\mu\nu} + k^{4}_{5}S_{\mu\nu} -E_{\mu\nu}.
\end{equation}
In the above equation $S_{\mu\nu}$ is the local quadratic energy-momentum correction
\begin{equation}
S_{\mu\nu}=\frac{1}{12}TT_{\mu\nu} -\frac{1}{4}{T_{\mu}}^{\alpha}T_{\nu\alpha}+ \frac{1}{24}g_{\mu\nu}(3T^{\alpha\beta}T_{\alpha\beta}-T^{2}),
\end{equation}
and $E_{\mu\nu}$  is the nonlocal effect from the bulk free gravitational field transmitted projection of the
bulk Weyl tensor $C_{abde}$,
\begin{equation}
E_{ab}=C_{aebd}n^{e}n^{d}, E_{ab} \rightarrow E_{\mu\nu} \delta^{\mu}_{a}\delta^{\nu}_{b}~~ \text{as}~~ \chi \rightarrow0.
\end{equation}
The four-dimensional cosmological constant, $\Lambda$, and the coupling constant $k_{4}$, are given by
\begin{equation}
\Lambda=\frac{k^{2}_{5}}{2}\left(\Lambda_{5}+\frac{k^{2}_{5}}{6}\right), ~~k^{2}_{4}=\frac{k^{4}_{5}}{6}.
\end{equation}

The Einstein equation in the bulk (Codazzi equation), also implies the conservation of the energy-momentum tensor
of the matter on the brane: $D_{\mu}T^{\mu}_{\nu}=0$. Moreover, the contracted Bianchi identities on the brane imply that the projected Weyl tensor should obey the following constraint: $D_{\mu}E^{\mu}_{\nu}=k^{4}_{5}D_{\mu}S^{\mu}_{\nu}$.
Finally,  Eq.(3) and the latter above give the complete set of field equations for the brane gravitational field.

For any non dissipate matter field the general form of the brane energy-momentum tensor can be covariantly written as
\begin{equation}
T_{\mu\nu}=\rho u_{\mu} u_{\nu} + p h_{\mu\nu}.
\end{equation}
The  decomposition is irreducible for any chosen four vector $u^{\mu}$. Here $\rho$ and $p$ are the energy density
and the isotropic pressure, and $h_{\mu\nu}=g_{\mu\nu} + u_{\mu}u_{\nu}$ projects orthogonal to $u^{\mu}$. The symmetric properties of $E_{\mu\nu}$ imply that, in general, we can decompose it irreducibly with respect to a chosen four-velocity field $u^{\mu}$ as
\begin{equation}
\label{Etensor}
E_{\mu\nu}=-k^{4}\left[U\left( u_{\mu}u_{\nu}+\frac{1}{3} h_{\mu\nu}\right)\right],
\end{equation}
where $k=k_{5}/k_{4}$. In Eq.(\ref{Etensor}) $U$ is the effective nonlocal energy density on the brane arising from free gravitational field in the bulk.


\section{Brane World-Model}


We consider a brane universe with the induced metric given by the Friedmann-Roberston-Walker(FRW) space-time. In the following, we shall explore the evolution of a cosmological brane filled with a tachyon  field $\varphi$  or a quintessence field $\phi$  and a perfect fluid. These fields have an energy-momentum tensor on the brane given by $T_{\mu\nu}=\ro_{a}u_{\mu} u_{\nu} + p_{a}h_{\mu\nu}$, where the label $a$ denotes the tachyon or quintessence {respectively}. For a homogeneous tachyon field the energy density and pressure are given by

\be
\n{rpt}
\rho_\va={U(\va)\over\sqrt{1-\dot\va^2}},\qquad p_\va=-{U(\va)\sqrt{1-\dot\va^2}}.
\ee
while for the quintessence the density energy and pressure take the following form:
\be
\n{rpq}
\rho_{\phi}=\frac{\dot{\phi}^2}{2}+ V(\phi),\qquad p_{\phi}=\frac{\dot{\phi}^2}{2}- V(\phi),
\ee
where $U(\va)$ and $V(\phi)$ are the potentials. We suppose that the barotropic components, radiation, dust and cosmic string  have equation of states, $p_{m}=(\gamma_m-1)\rho_m$ where the barotropic indices are $\ga_m=4/3,1$ and $2/3$ respectively. They exchange energy with tachyon or quintessence fields only gravitationally, so that, each one of the matter and field components satisfies separate equations of conservation.
Under this scheme the dynamics on the brane is completely determined by the following set of equations
 \ben
 3\left(H^{2}+\frac{k}{a^2}\right)=\Lambda+\rho+ \frac{3}{\lambda^2}(\rho^2 + 12{\cal U})\label{fm}
,\\
\dot\rho_{a}+3H(\rho_{a}+p_{a})=0,\label{cphi1}
\\
\dot\rho_{m}+3H\rho_{m}\gamma_{m}=0,\label{cmat}
\\
\frac{36}{\lambda^2}{\cal U}=-\frac{\Gamma^2}{a^4},\label{det}
 \een
where $k=-1,0,1$ is the spatial curvature, and $\rho=\rho_{m}+ \rho_{a}$ is the total energy density. In the absence of anisotropic stress, the nonlocal energy density  $\cal U$, containing the effects from the bulk, takes the form of dark radiation  with its {behavior}
given by Eq. (\ref{det}).
As a final remark, one can see that the evolution of the early universe can be separated into eras. In the high  energy regime, $3\rho^2/\lambda^2\gg\rho$, the quadratic density term in Eq. (\ref{fm}) becomes dominant and we get an unconventional expansion law for the universe  $H^2\approx\rho^{2}/\lambda^{2}$. However, in the low energy regime, $3\rho^2/\lambda^2\ll\rho$, the linear density term in Eq. (\ref{fm}) dominates and we recover the standard Friedmann equation in four dimension $3H^{2}\approx\rho$.
Below, we develop a simple algorithm for tachyon and quintessence fields that allows us to describe the evolution of the Universe from radiation to cosmic string eras. In addition, it will be used to find exact {solutions} for the modified Friedmann equation on the brane.


\subsection{Tachyon case (TC)}


Because of the recent interest of considering a tachyon field component in cosmology we investigate a brane-world universe where the field localized on the brane is given by a single tachyon \cite{tachyonherrera}, \cite{egbtachyonsami}, \cite{perinflabrane}.
We wish to extract information from the modified Einstein equations assuming that the energy density depends  only on  the scalar factor $\rho_{\varphi}=\rho_{\varphi}(a)$. Inserting it in the conservation equation (\ref{cphi1}) we obtain
 \begin{equation}
3\dot{\varphi}^2=-a\frac{\rho'_{\varphi}}{\rho_{\varphi}},\label{kedea1}
 \end{equation}
where the prime denotes a differentiation with respect to the scale factor. Combining Eqs. (\ref{rpt}), (\ref{fm}) and (\ref{kedea1}) we may write the potential and the tachyon field
 \begin{equation}
U(\varphi)=\Big[\rho^{2}_{\varphi} + \frac{a}{3}\rho'_{\varphi}\rho_{\varphi}\Big]^{1/2},\label{Vdea1}
 \end{equation}
\begin{equation}
\label{phit}
\varphi=\pm\int{\left[\frac{-{\rho'}_{\varphi}}{a\rho_{\varphi}(\Lambda-\frac{3k}{a^2} +\rho+ \frac{3}{\lambda^2}\rho^2 - \frac{\Gamma^2}{a^4})}\right]^{1/2}}da,
\end{equation}
as functions of the scale factor. Integrating the last equation leads to  $\varphi(a)$ and inverting it gives $a(\varphi)$. Finally, by using (\ref{Vdea1}) it follows the potential $U(a(\varphi))$. Thus the above procedure induces a model where the tachyon field is driven by this potential with an exact scale factor produced by the desired energy density $\rho_{\varphi}(a)$.


\subsection{Quintessence case (QC)}


Following the same steps above we write the energy density of the
quintessence scalar field as a function of scale factor
$\rho_{\phi}=\rho_{\phi}(a)$.
Inserting it in the conservation equation (\ref {cphi1}), we obtain

\begin{equation}
3 \dot{\phi}^2=-a\rho'_{\phi}.
\label{kedea2}
\end{equation}

>From this equation we are able to write the potential energy as a function
of the scale factor,
\begin{equation}
V(a)=\rho_{\phi} +\frac{a}{6}\rho'_{\phi}.\label{Vdea2}
 \end{equation}
Employing   Eqs. (\ref{fm}) and (\ref{kedea2}) we find the quintessence field $\phi(a)$
\begin{equation}
\label{phiq}
\phi=\pm\int{\left[\frac{-{\rho'}_{\phi}}{a(\Lambda-\frac{3k}{a^2} +\rho+ \frac{3}{\lambda^2}\rho^2 - \frac{\Gamma^2}{a^4})}\right]^{1/2}}da.
\end{equation}
Replacing $\rho_\phi(a)$ in the last equation, it gives $\phi(a)$. Inverting it gives $a(\phi)$ and by using (\ref{Vdea2}), it follows $V(a(\phi))$. Thus the procedure determines $V(\phi)$ and defines a model with an exact solution on the brane with the desired energy density $\rho_{\phi}(a)$. In the next section we shall use the algorithm for tachyon and quintessence field to depict different cosmological eras.


\section{Description of the radiation, matter and cosmic string eras}


We investigate a spatially flat FRW universe with no cosmological constant and nonvanishing dark radiation energy density, filled with a tachyon or quintessence field and a barotropic component. In what follows, we shall examine the cosmological evolution of the Universe from the radiation to the cosmic string eras by using the model developed in the  Sect. III.  In this case the Friedmann equation reads
\ben
\n{cero}
3H^{2}=\rho+ \frac{3}{\lambda^2}\rho^2-\frac{\Gamma^2}{a^4}
 \een
with $\rho=\rho_{m}+ \rho_{a}$. To solve the nonlinear Eqs. {(\ref{cphi1}), (\ref{cmat}) and (\ref{cero})} we assume that both the energy densities of the tachyon and quintessence fields have an inverse power law  dependence with the scale  factor $\rho_{\varphi}=\rho_{\varphi0}a^{-s}$, $\rho_{\phi}=\rho_{\phi0}a^{-q}$, and the barotropic index satisfies the following inequalities $s,q < 3\gamma_m$. The nonlocal effects from the bulk will be important in the first radiation era and neglected in the remaining   ones.


\subsection{Radiation dominated era}


In the TC and QC, we split the radiation dominated era, with $\rho_m=\ro_r=\rho_{r0}/a^4$, into two stages {corresponding to} the high and low energy regimes. These regimes are separated by the critical energy density $\ro_c=\lambda^2/3$ where the linear and quadratic terms in the Friedmann equation are comparable.  For the TC, in the higher energy regime, $\rho\gg\rho_c$, the tachyon field  and the potential have the following behaviors
\begin{equation} \label{VT1}
\varphi_r\approx \frac{\lambda}{4\rho_{r0}}\sqrt{\frac{s}{3}} a^{4},\qquad U_r\approx U_{r0}~\varphi^{-s/4}.
\end{equation}
For the QC, we find that the quientessence field is given by
\begin{equation} \label{phia1}
\phi_r\approx\frac{2\lambda}{\left(8-q\right)\rho_{r0}}
\sqrt{\frac{q\rho_{\phi 0}}{3}}a^{4-q/2},
\end{equation}
and the potential reads
\begin{equation} \label{Vphi1}
V_r\approx V_{r0} ~\phi^{-2q/\left(8-q\right)}.
\end{equation}
In this way we recover the results obtained in Ref. \cite{Maeda}, and  the proposal of a tracking potential arises as a consequence of the domination assumptions.

As long as the Universe expands the energy density decreases and there exists a critical cosmological time where the quadratic and linear terms become equal. After that the quadratic term becomes subdominant, it begins the low regime $\ro<\ro_c$, and the conventional cosmology is recovered. In Ref. \cite{Caroll} it was shown that nucleosynthesis is not restricted to the
conventional radiation domination era and could have begun earlier, during the
late brane radiation domination era. Then, assuming that $T_{c}>T_{NS}\approx 1$ MeV we
recover the results obtained in Ref. \cite{Maeda}, \cite{Huey}.

When the Universe enters in the second radiation era  the effects of the brane becomes
negligible. In this cosmological stage, for the TC the tachyon field and the potential become
\begin{equation} \label{VT2}
\varphi_r\approx \frac{1}{2}\sqrt{\frac{s}{\rho_{r0}}} a^{2},\qquad U_r\approx U_{r0}\varphi^{-s/2},
\end{equation}
whereas for the QC, the quintessence field and the potential energy exhibit the following behaviors:
\begin{equation} \label{phia2}
\phi_r\approx \frac{2}{q-4}\sqrt{\frac{q}{r}}a^{2-q/2}, \qquad V_r\approx V_{r0}\phi^{-2q/(4-q)},
\end{equation}
where $r=\ro_{r0}/\ro_{\phi 0}$.
Solving Eqs. (\ref{cmat}) and (\ref{cero}) for radiation eras with $\ga_m=4/3$ and $\ro_m=\ro_r=\ro_{r0}/a^4$, we obtain the scale factor
\be
\n{ar}
a^{4}_r\approx\frac{4(\ro_{r0} -\Gamma^{2})}{3}\,t^{2}+ \frac{4\ro_{ro}}{\lambda}\,t.
 \ee
Because of the contribution of the nonlocal dark radiation $-\Gamma^{2}/a^{4}$ the coefficient of the quadratic term has not definite sign showing the relevance of the bulk. Let us examine the possible behaviors of the brane-world universe dominated by radiation. For $\ro_{r0}>\Gamma^{2}$ the singular solution (\ref{ar}) represents a universe which interpolates between $a\approx t^{1/4}$ in the high energy regime and $a\propto t^{1/2}$ in the low energy regime. However, in the case $\ro_{r0} <\Gamma^{2}$ the nonlocal dark radiation component makes it possible for the  Universe to collapse in a big crunch at   $t_{c}=3\ro_{r0}/\la (\Gamma^{2} - \ro_{r0})$. The Universe has a finite time span and the scale factor has a maximum at $a^4_{max}=3\ro^{2}_{r0}/\la^{2}(\Gamma^{2} - \ro_{r0})$, where the total energy density vanishes and the acceleration $\ddot{a}$ is negative. In the particular case $\ro_{r0} =\Gamma^{2}$,  the scale factor $a\approx(\frac{4\ro_{ro}}{\lambda})^{1/4}t^{1/4}$ has an initial singularity.


\subsection{Cold dark mater (CDM) dominated era}


As soon as the radiation thermalizes and equilibrates with the baryonic matter as well as  with the nonrelativistic
nonbaryonic cold dark matter (whose energy-momentum tensor is dustlike in the first approximation $0<p_{DM}\ll\rho_{DM}$) the Universe is dominated by all these contributions. Actually, we  {have included} both baryonic and dark matter energy densities in the same term, $\ro_{_{CDM}}=\ro_{_{CDM0}}/a^3$. Assuming  that $s,q<3$, we get in the TC, the following expressions for the tachyon field and its potential energy

\begin{equation} \label{VT3}
\varphi_{_{CDM}}  \approx \frac{2}{3} \sqrt{\frac{s}{\rho_{_{CDM0}}}} a^{3/2},\quad
U_{_{CDM}}  \approx U_{_{CDM0}} \varphi^{-2s/3},
\end{equation}
while  for the QC, we have
\be \label{phia3}
\phi_{_{CDM}}\approx \frac{2}{q-3}\sqrt{\frac{q}{r}}a^{(3-q)/2}, \quad
V_{_{CDM}}\approx V_{_{CDM0}}\phi^{-2q/(3-q)},
\ee
with $r=\ro_{_{CDM0}}/\ro_{\phi 0}$.

Typically, in this CDM era the effects of the brane diminish and the contribution of the nonlocal dark radiation term $-\Gamma^{2}/a^{4}$ may be neglected. Hence,  by solving Eqs. (\ref{cmat})-(\ref{cero}) one finds that the approximate scale factor is given by $a\approx (3\rho_{_{CDM0}}/4)^{1/3}\,t^{2/3}$ and the Universe behaves as if it was dominated by matter.


\subsection{Cosmic string dominated era}


When the dark matter dominated era ends and before the Universe begins to feel the dark energy component,
which is the probable responsible for its currents acceleration, the Universe is dominated by cosmic string networks. This intermediate regime, that is, the transition between the CDM regime with $\rho_{_{CDM}}=\ro_{_{CDM0}} a^{-3}$ (non accelerated stage) and the dark energy era (accelerated phase), is properly described by the cosmological cosmic string energy density $\rho_{st}=\rho_{st0}a^{-2}$. Then, in the cosmic string era, the tachyon field and potential in the TC take the following form

\begin{equation} \label{VT4}
\varphi_{st}\approx \sqrt{\frac{s}{\rho_{st0}}} a,\quad U_{st}\approx U_{st0}~\varphi^{-s}.
\end{equation}
and
\begin{equation} \label{phia4}
\phi_{st}\approx \frac{2}{q-2}\sqrt{\frac{q}{r}} a^{\frac{2-q}{2}},\quad V_{st}\approx V_{st0}~\phi^{-2q/(2-q)},
\end{equation}
for the QC with $r=\rho_{st0}/\rho_{\phi0}$.

In this era, discarding the effects of the brane and the contribution of the nonlocal dark radiation term, the cosmic string networks density makes the scale factor go linear with the cosmic time and has the form $a\approx (\ro_{st0}/3)^{1/2}\,t$.

In conclusion, we have found that in every  stage of the Universe, the corresponding tachyon and quintessence fields are driven by inverse power law potentials. It is  important to remark that  {these kinds of potentials} are very interesting for several reasons. For instance, they arise in supersymmetric  {condensate} models of QCD \cite{Binetruy-Brax} and can in principle act as a source of quintessence for a brane-world \cite{Maeda}.


\section{ Generating exacts brane-worlds solutions}



\subsection{Quintessence solutions}


Let us conceive the energy density of the quintessence field $\ro_\phi$ as a mixture of vacuum energy density $\ro_v=\al$, with $\al>0$, plus a cosmic string network density interpreted as a nonrelativistic matter whose barotropic index is $\gamma=2/3$, $\ro_s=\ro_{\phi 0}/a^2$ \cite{cosmicstring}, with $\ro_{\phi 0}>0$, so
\be
\n{Ro}
\ro_\phi = \alpha +\frac{\ro_{\phi 0}}{a^2}.
\ee
With the idea of relating our finding with  {the} results obtained in \cite{Ellis}, we shall be restricted from now on  to the case of $\rho_m=0$.

\vskip .5cm


\no {\it 1.} \quad $\Gamma^{2}=\frac{3\ro^{2}_{\phi 0}}{\la^2}$, \,\,$\ro_{\phi 0}\left(1+\frac{6\al}{\la^2}\right)>3k$,  and \,\,$\Lambda>-\al\left(1+\frac{6\al}{\la^2}\right)$.


\vskip .5cm

Integrating Eq. (\ref{fm}), the scale factor becomes
\be
\n{a1a}
a(t)=\frac{\sqrt{\nu}}{\sqrt{3}\,\omega}\,\sinh{\omega t},
\ee
where $\nu=\ro_{\phi 0}(1+6\al/\la^2)-3k$. It represents an expanding singular universe with a final de Sitter scenario.  By following Ref. \cite{Ellis} we work with the constants
\ben
\omega^2=\frac{1}{3}\left[\Lambda+\al\left(1+3\frac{\al}{\lambda^2}\right)\right], \quad \Gamma^2=\frac{27}{p^2}\left(k+
\frac{\nu}{3}\right)^2,\\
p^2=\lambda^2+36\left(\omega^2-\frac{\Lambda}{3}\right).\,\,\,\,\,\,\,\,\,\,\,\,\,\,\,\,\,\,\,\,\,\,\,\,\,\,\,\,\,\,\,\,\,\,\,\,\,\,\,\,\,\,\,\,\,\,\,\,\,\,\,\,\,\,\,\,\,\,\,\,\,\,\,\,\,\,
\een
By combining Eqs. (\ref{kedea2}), (\ref{Ro}), (\ref{a1a}) and integrating, we have the field as a function of the cosmic time
\be
\n{f1a}
\Delta\phi=\pm\,\sqrt{{\frac{2\lambda}{p}\left(1+\frac{3k}{\nu}\right)}}\ln\left|\tanh{\frac{\omega\,t}{2}}\right|,
\ee
Now, by making the composition of the scale factor (\ref{a1a}) with the scalar field (\ref{f1a}), we obtain $a=a(\phi)$. Then, from Eqs. (\ref{Vdea2}) and (\ref{Ro}), we get the potential
\ben
\nonumber
\n{v1a}
V(\phi)=\frac{\lambda^2}{6}\left(\frac{p}{\lambda}-1\right)+\frac{2\lambda\omega^2}{p} \left(1+\frac{3k}{\nu}\right)\\
\times\sinh^2\left[\sqrt{ {\frac{p}{2\lambda\left(1+\frac{3k}{\nu}\right)}}} \,\Delta\phi\right].\qquad
\een
At the initial singularity the scalar field and the potential diverge but in the limit $t\rightarrow \infty$ the scalar field vanishes and the potential reaches its minimum value. This means that the exact solution is stable. For $k=\Lambda=0$ and $\nu=3\omega^2$ the solution mentioned above coincides with the solution reported in Ref. \cite{Ellis}.

\vskip .5cm


\no {\it 2.} \quad $\Gamma^{2}=\frac{3\ro^{2}_{\phi 0}}{\la^2}$, \,\, $\ro_{\phi 0}\left(1+\frac{6\al}{\la^2}\right)>3k$,  and \,\,$\Lambda<-\al\left(1+\frac{6\al}{\la^2}\right)$.


\vskip .5cm

Now, the solution of  Eq. (\ref{fm}) represents a singular universe with a finite time span
\be
\n{a1c}
a(t)=\frac{\sqrt{\nu}}{\sqrt{3}\,\omega}\,\sin{\omega t},
\ee
which begins at $t=0$ and ends in a ``big-crunch" at $ t_{BC}=\pi/\omega$. Also, from (\ref{kedea2}), (\ref{Ro}) and (\ref{a1c}), we get the dependence of the scalar field with the cosmic time and the potential (\ref{Vdea2}) as a function of the scalar field, namely,
 \ben
 \n{f1c}
 \Delta\phi=\pm\sqrt{\frac{2\lambda}{p}\left(1+\frac{3k}{\nu}\right)}
 \ln\left|\tan\frac{\omega t}{2}\right|,\\
 \nonumber
V(\phi)=\frac{\lambda^2}{6}\left(\frac{p}{\lambda}-1\right)+\frac{2\lambda\omega^2}{p}\left(1+\frac{3k}{\nu}\right)\\
 \n{v1b}
\times\cosh^2{\left[\sqrt{\frac{p}{2\lambda\left(1+\frac{3k}{\nu}\right)}}\,\Delta\phi\right]}.\qquad
 \een
Note that the potential takes its minimum value at the maximum value of the scale factor and diverges at the initial and final singularities.


\subsection{Tachyon solutions}


Using the method introduced in Sec. III.A for the energy density $\rho_{\varphi}=\rho_{\va 0} a^{-2}$,  we shall find new solutions for the tachyonic field localized on the brane.

\vskip .5cm

\no {\it 1.}\qquad $\Gamma^{2}=\frac{3\ro^{2}_{\va 0}}{\la^2}$, \quad $\ro_{\va 0}>3k$, and  \quad $\Lambda>0$.

\vskip .5cm

By solving Eqs. (\ref{fm}) and (\ref{phit}), we obtain the scale factor and the tachyon field

\ben
\n{a1t}
a(t)=\sqrt{\frac{\nu}{\Lambda}}\sinh \sqrt{\frac{\Lambda}{3}}t, \\
\n{tach1}
\Delta\varphi= \sqrt{\frac{2}{\Lambda}}{\rm arcsinh}\sqrt{\frac{\Lambda}{\nu}}a.
\een

where $\nu=\ro_{0\varphi}-3k$. By considering $a=a(\varphi)$ and $\ro_\va$ in Eq. (\ref{Vdea1}),
we find the potential $U(\varphi)$ for the tachyon field
\be
\n{Vt1}
U(\varphi)= \frac{ \Lambda {\ro_{0\varphi}}  }{\sqrt{3}\,\nu\sinh^2{m \Delta\varphi}}
\ee

where $m=(\Lambda/2)^{1/2}$ and $\Delta\varphi=\varphi-\varphi_{0}$. Let us examine the  {behavior} of the scale factor  and the potential  {energy}. At the initial singularity, the potential energy of the tachyon field blows up but  for large cosmological time, the scale factor has a final de  Sitter stage $a\propto e^{\sqrt{\frac{\Lambda}{3}}t}$ and $U(\va)$   becomes an exponential potential $U\approx 4\Lambda e^{-\sqrt{\Lambda}\Delta\varphi}/{\sqrt{3}\,\nu}$ so the solution is stable.

\vskip .5cm

\no {\it 2.}
\qquad $\Gamma^{2}=\frac{3\ro^{2}_{\va 0}}{\la^2}$, \quad $\ro_{\va 0} >3k$, and \quad $\Lambda<0$.

\vskip .5cm

After using Eqs. (\ref{fm}) and (\ref{phit}), the scale factor and tachyon field are given by

\ben
\n{a2t}
a(t)=\sqrt{\frac{\nu}{-\Lambda}}\sin\sqrt{\frac{-\Lambda}{3}}t,   \\
\label{tach2}
\Delta\varphi= -\sqrt{\frac{2}{-\Lambda}}\arcsin\sqrt{\frac{-\Lambda}{\nu}}a.
\een
 {By inserting Eq.} (\ref{tach2}) into (\ref{Vdea1}), we get the potential for the tachyon  {field}
\be
\n{Vt2}
U(\varphi)= \frac{-\Lambda{\ro_{0\varphi}} }{\sqrt{3}\,\nu\sin^{2}(-\Lambda/2)^{1/2}\Delta\varphi}.
\ee
In this case, the Universe has a finite time span and the tachyon field vanishes at the initial and final singularities while the potential diverges as $U\approx a^{-2}$. With the help of Eq.(\ref{cphi1}) and the tachyon energy $\rho_{\varphi}=\rho_{\va 0} a^{-2}$ we obtain the relation $\rho_{\varphi}+ p_{\varphi}=\frac{2}{3}\rho_{\varphi}\geq 0$ which shows that the tachyon solutions satisfy the weak energy condition ($\rho\geq0$, $\rho+p\geq 0$) and null energy condition( $\rho+p\geq 0$) conditions. As a final remark, we desire to stress that a very large set of solutions can be found extending the previous proposal on the energy density. In a forthcoming paper, starting from the energy density $\rho(a)=\al + \ro_{0}a^{-n}$ with $n>0$, we will find several types of solutions of physical interest.


\section{Conclusion}


To sum up, we have presented a new algorithm for solving the brane-world
field equations  on  FRW  backgrounds with  different types of sources.
Basically, we work with a single tachyon, a classical minimally coupled
scalar field and barotropic matter which are confined to the brane, embedded
in five-dimensional Einstein gravity. Furthermore, we find it is possible to mimic with a tachyon (or quintessence) field the  different cosmological stages, starting from  {a} first radiation era to  {a} universe dominated by a cosmic string  networks era. In general, we have found that the potential for the tachyon (or quintessence) field is given by an inverse power law $U=U_{0}~\varphi^{-|n|}$. Also, our procedure seems to match quite well with the phenomenology of  brane-world quintessence analyzed by Maeda \cite{Maeda} and Huey and Lidsey \cite{Huey}. Additionally,  we have found that due to the nonlocal dark radiation term $-\Gamma^{2}/a^{4}$ the Universe could exhibit a big crunch.  Besides,  with  this algorithm we reproduce and generalize the solutions found in \cite{Ellis}, \cite{Sami}. Later, we extend the previous analysis and report exact  solutions of the full brane field equations with tachyon and/or quintessence sources.


\acknowledgments


The authors acknowledge the partial
support under Project No. 24/07 of the  agreement SECYT (Argentina) and CAPES 117/07 (Brazil).
LPC thanks the University of Buenos Aires for partial support under
Project No. X224, and the Consejo Nacional de Investigaciones
Cient\'{\i}ficas y T\'ecnicas under Project No. 5169. GMK acknowledges the support by
Conselho Nacional de Desenvolvimento Cient\'{\i}fico e Tecnol\'ogico (CNPq). MGR is supported by  the Consejo Nacional de Investigaciones Cient\'{\i}ficas y T\'ecnicas and acknowledges the hospitality of the Physics Department
of Universidade Federal do Paran\'{a}, (Curitiba) where a part of this
work was done.




\begin{thebibliography}{99}












\bibitem{hw1} P. Ho\v rava and E. Witten, {\em Nucl. Phys.} {\bf B460}
    (1996) 506.

\bibitem{hw2} P. Ho\v rava and E. Witten, {\em Nucl. Phys.} {\bf B475}
(1996) 94.

\bibitem{Gogberashvili:1999ad}
M.~Gogberashvili, Int. J. Mod. Phys. D {\bf 11}, 1639 (2002), 
[arXiv:hep-ph/9908347].

\bibitem{Ra1}
L.~Randall and R.~Sundrum, Phys. Rev. Lett. {\bf 83}, 4690-4693 (1999)
[arXiv:hep-th/9906064].

\bibitem{Cohen:1999ia}
A.G.~Cohen and D.B.~Kaplan, Phys. Lett. B {\bf 470}, 52 (1999),
[arXiv:hep-th/9910132].

\bibitem{Gregory:1999gv}
R.~Gregory, Phys. Rev. Lett. {\bf 84}, 2564 (2000),
[arXiv:hep-th/9911015].

\bibitem{Visser:1985qm}
M.~Visser,Phys. Lett. {\bf 159B}, 22 (1985),
[arXiv:hep-th/9910093].

\bibitem{Ra2}
L. Randall and R. Sundrum, Phys. Rev. Lett. {\bf 83}, 3370-3373 (1999), [arXiv:hep-th/9905221].



\bibitem{Sh1}
T. Shiromizu, K. Maeda, and M. Sasaki, Phys.Rev. D {\bf 62}, 024012 (2000), [arXiv:gr-qc/9910076].

\bibitem{Cs2}
C. Cs\'{a}ki, M. Graesser, L. Randall, and J. Terning, [arXiv:hep-th/9911406]

\bibitem{rs2}
N. Arkani-Hamed, S. Dimopoulos, G. Dvali and N. Kaloper,
Phys. Rev. Lett. {\bf 84} (2000) 586.


\bibitem{gt}
J. Garriga and T. Tanaka, [arXiv:hep-th/9911055].

\bibitem{gkr}
S.B. Giddings, E. Katz and L. Randall,[arXiv: hep-th/0002091].


\bibitem{ADD}
N. Arkani-Hamed, S. Dimopoulos and G. Dvali, Phys. Lett. B 429, 263 (1998).
I. Antoniadis, N. Arkani-Hamed, S. Dimopoulos and G. Dvali,
Phys. Lett. B{436}, 257 (1998).

\bibitem{cftbrane}

A. Fayyazuddin and M. Spalinski, Nucl. Phys. {\bf B} 535, 219
(1998), [arXiv:hep-th/9805096]; O. Aharony, A. Fayyazuddin,
and J. Maldacena, J. High Energy Phys. {\bf 07}, 013 (1998),
[arXiv:hep-th/9806159].

\bibitem{egb5d}
D. Lovelock, J. Math. Phys. 12, 498 (1971); N. Deruelle and J. Madore, Mod. Phys. Lett. {\bf A}1, 237 (1986); N. Deruelle and L. Farina-Busto, Phys. Rev. {\bf D} 41, 3696 (1990).

\bibitem{charmoux}
C.Charmousis and J-F. Dufaux, Class.Quant.Grav. {\bf19},4671(2002),[arXiv:hep-th/0202107];
 J.E. Lidsey and N.J. Nunes, Phys. Rev. D {\bf67}, 103510 (2003).

\bibitem{kmaeda}
Kei-ichi Maeda, Takashi Torii, Phys.Rev. D {\bf69}, 024002 (2004), [arXiv:hep-th/0309152].

\bibitem{dufaux}
J.-F. Dufaux, J. Lidsey, R. Maartens, M. Sami, Phys.Rev. D {\bf70}, 083525 (2004),[arXiv:hep-th/0404161].

\bibitem{egbotros}
Shin'ichi Nojiri, Sergei D. Odintsov,J. High Energy
Phys. {\bf 07} (2000) 049, [arXiv:hep-th/0006232]; B. Abdesselam and N. Mohammedi, Phys. Rev. D {\bf65}, 084018 (2002); J. E. Lidsey, Shin'ichi Nojiri, S. D. Odintsov, J. High Energy Phys. 06 (2002)026[arXiv:hep-th/0202198]; S.C. Devis, Phys. Rev. D {\bf67}, 024030 (2003)[arXiv:hep-th/0208205]; E. Gravanis and S. Willison,Phys. Lett. B 562, 118 (2003) [arXiv:hep-th/0209076]; S. Nojiri,
S.Odintsov and S. Ogushi, Phys. Rev. D {\bf65} (2002) 023521; J. P. Gregory, A. Padilla, [arXiv:hep-th/0304250]; J. E. Kim and H. M. Lee, Nucl. Phys.
B{\bf602}, 346 (2001); B{\bf619}, 763(E) (2001);N. Deruelle
and C. Germani, Nuovo Cimento Soc. Ital. Fis. B {\bf118},
977 (2003),[arXiv:gr-qc/0306116]; Nathalie Deruelle and John Madore, [arXiv:grqc/
0305004]; Kyong Hee Kim and Yun Soo Myung, J.
Cosmol. Astropart. Phys. {\bf12} (2004) 004; Kyong Hee Kim
and Yun Soo Myung, Int. J. Mod. Phys. D {\bf14}, 1813 (2005), [arXiv:astro-ph/0408278];
Yun Soo Myung, Phys. Lett. B {\bf601}, 1 (2004); J. E. Lidsey,
S. Nojiri, and S. D. Odintsov, J. High Energy Phys. {\bf06}
(2002) 026. 


\bibitem{dgp}
G. R. Dvali, G. Gabadadze and M. Porrati, Phys. Lett. {\bf B} 485 (2000) 208 [arXiv:hep-th/0005016].

\bibitem{dgpcosmo}
C. Deffayet, G. R. Dvali and G. Gabadadze, Phys. Rev. D {\bf65}, 044023 (2002)[arXiv:astro-ph/0105068].

\bibitem{sentachyon}
A. Sen, Mod. Phys. Lett. A {\bf17}, 1797 (2002).

\bibitem{samit}
M. Sami , P. Chingangbam and T. Qureshi, Phys. Rev. D {\bf66}, 043530 (2002).


\bibitem{tachyongibbons}
G. W. Gibbons, Phys. Lett. B {\bf 537}, 1 (2002).

\bibitem{extendedtachyon}
L. P. Chimento, Phys.Rev.D {\bf69},123517 (2004), [arXiv:astro-ph/0311613].

\bibitem{bruselas}
M. Fairbairn and M. H.G. Tytgat,  Phys.Lett. B {\bf 546}, 1-7 (2002), [arXiv:hep-th/0204070].

\bibitem{ruthtachyon}
J.M. Aguirregabiria, Ruth Lazkoz,  Mod.Phys.Lett.A {\bf19} 927-930,2004,  [arXiv:gr-qc/0402060].

\bibitem{Paultachyoninflation}
B.C. Paul, Dilip Paul,  [arXiv:0708.0897].

\bibitem{tachyonpl}

Yongli Ping, Lixin Xu, Hongya Liu, Ying Shao [arXiv:hep-th/08010268].


\bibitem{tachyonherrera}
Ramon Herrera, [arXiv:gr-qc/0810.1074].

\bibitem{MaartensExact}

R.Maartens, D.Wands, B.A.D Basselt and I.P.C. Heard. Phys. Rev. D. {\bf62}, 041301 (2000).

\bibitem{PaulExact}

B.C.Paul,  Phys.Rev. D {\bf68}, 127501 (2003), [arXiv:hep-th/0309205v1].

\bibitem{HawkinsLidsey}

R. M. Hawkins and J. E. Lidsey, Phys.Rev.D {\bf63}, 041301 (2001), [arXiv: gr-qc/0011060].

\bibitem{Maartens} R. Maartens,
{\it Phys. Rev.} D {\bf 62}, 084023 (2000), [arXiv:hep-th/0004166];
{\em Geometry and Dynamics of the Brane-World}, [arXiv:gr-qc/0101059].

\bibitem{MaWa}
K.Maeda and D.Wands, Phys. Rev. D {\bf 62}, 124009 (2000).


\bibitem{egbtachyonsami}
M. Sami, N. Savchenko and A. Toporensky,  Phys. Rev .D {\bf70}, 123528 (2004), [arXiv:hep-th/0408140].

\bibitem{perinflabrane}
L. Leblond, S. Shandera,J. Cosmol. Astropart. Phys.
01 (2007) 009, [arXiv:hep-th/0610321].

\bibitem{Maeda}
K. Maeda,  Phys.Rev. D {\bf64}, 123525 (2001), [arXiv: astro-ph/0012313].

\bibitem{Caroll}
S.M. Caroll and M. Kaplinghat, [arXiv:astro-ph/0108002].
\bibitem{Huey}
G. Huey and J. E. Lidsey,  Phys.Lett.B {\bf514}, 217-225 (2001), [arXiv:astro-ph/0104006].


\bibitem{Binetruy-Brax}
P. Binetruy, Phys.Rev. D {\bf60}, 063502 (1999); P. Brax and J. Martin,  Phys.Rev. D {\bf61}, 103502 (2000).

\bibitem{cosmicstring}
A. Vilenkin, Phys. Rev. Lett. {\bf 53}, 1016 (1984); Phys. Rep.
{\bf 121}, 263 (1985); A. A. Soleng, Gen. Relativ. Gravit. {\bf 27},
367 (1995); M. P. Dabrowski and J. Stelmach, Astron. J.
{\bf 97}, 978 (1989).

\bibitem{Ellis}
D. Solomon, P. Dunsby and G. Ellis, [arXiv:gr-qc/0102016].

\bibitem{Sami}
M. Sami,Gravitation Cosmol. {\bf7}, 228 (2001),[arXiv:gr-qc/0105052].

\end{thebibliography}
\end{document}